\documentclass[12pt]{article}

  \usepackage{graphicx}
  \usepackage{epsfig}
  \usepackage{amssymb}
  \usepackage{subfigure}
    \usepackage{feynmf}
\evensidemargin - 1.truecm
\begin{document}

\newcommand{\be}{\begin{equation}}
\newcommand{\ee}{\end{equation}}
\newcommand{\nn}{\nonumber}
\newcommand{\bea}{\begin{eqnarray}}
\newcommand{\eea}{\end{eqnarray}}
\newcommand{\bfig}{\begin{figure}}
\newcommand{\efig}{\end{figure}}
\newcommand{\bc}{\begin{center}}
\newcommand{\ec}{\end{center}}
\newcommand{\bd}{\begin{displaymath}}
\newcommand{\ed}{\end{displaymath}}


\begin{fmffile}{Higgs}

\begin{titlepage}
\nopagebreak
{\flushright{
        \begin{minipage}{5cm}
         Rome1-1382/04 \\
         Freiburg-THEP 04/12 \\
         RM3-TH/04-16 \\
         IFUM-802/FT \\
        {\tt hep-ph/0407162}\\
        \end{minipage}        }

}
\renewcommand{\thefootnote}{\fnsymbol{footnote}}
\vskip 1.5cm
\begin{center}
\boldmath
{\Large\bf Master integrals for the two-loop light \\[3.mm] 
fermion contributions to $gg\rightarrow H$
and $H\rightarrow\gamma\gamma$
}\unboldmath
\vskip 1.cm
{\large  U.~Aglietti 
\footnote{Email: Ugo.Aglietti@roma1.infn.it} ,}
\vskip .2cm
{\it Dipartimento di Fisica, Universit\`a di Roma ``La Sapienza'' and
INFN, Sezione di Roma, P.le Aldo Moro~2, I-00185 Rome, Italy} 
\vskip .2cm
{\large  R.~Bonciani 
\footnote{This work was supported  by the European Union under
contract HPRN-CT-2000-00149}
\footnote{Email: Roberto.Bonciani@physik.uni-freiburg.de} ,}
\vskip .2cm
{\it Fakult\"at f\"ur Mathematik und Physik, 
Albert-Ludwigs-Universit\"at
Freiburg, \\ D-79104 Freiburg, Germany} 
\vskip .2cm
{\large G.~Degrassi\footnote{Email: degrassi@fis.uniroma3.it}},
\vskip .2cm
{\it Dipartimento di Fisica, Universit\`a di Roma Tre and 
INFN, Sezione di Roma III, \\ Via della Vasca Navale~84, I-00146 Rome, Italy} 
\vskip .2cm
{\large A.~Vicini\footnote{Email: Alessandro.Vicini@mi.infn.it}}
\vskip .2cm
{\it Dipartimento di Fisica, Universit\`a degli Studi di Milano and
INFN, Sezione di Milano,
Via Celoria 16, I--20133 Milano, Italy} 
\end{center}
\vskip 1.2cm

\begin{abstract}
We give the analytic expressions of the eight master integrals entering 
our previous computation of two-loop light fermion contributions to 
$gg\rightarrow H$ and $H\rightarrow\gamma\gamma$. The results are 
expressed in terms of generalized harmonic polylogarithms with maximum 
weight four included.

\vskip .4cm
{\it Key words}: Feynman diagrams, Multi-loop calculations

\end{abstract}
\vfill
\end{titlepage}    
%
%
%
%
In this note we give the analytic expressions of the eight
master integrals (MIs) entering our previous computation \cite{ABDV}
of two-loop light fermion contributions to 
\begin{equation}
\label{process1}
g+g\, \rightarrow \, H
\end{equation}
and 
\begin{equation}
\label{process2}
H \, \rightarrow \, \gamma+\gamma.
\end{equation}
As the MIs enter, in general, in various processes,
we believe that their publication is of general utility.
The Feynman diagrams for processes (\ref{process1}) and (\ref{process2}) 
are shown in Figs.~\ref{fig1} and \ref{fig2} respectively.
As is clearly seen, the amplitudes related to (\ref{process1}) 
are a subset of those for (\ref{process2}), so it is sufficient 
to consider only the latter process.

The reduction chain leading from the Feynman diagrams to the MIs
has been discussed briefly in \cite{ABDV} and in greater detail
in \cite{UgoRo1}, so we do not need to repeat the generalities 
but only the peculiarities to this case.
There are three independent topologies with six denominators 
shown in Fig.~\ref{fig3}. 
By shrinking one internal line, we obtain the five denominator
topologies listed in Fig.~\ref{fig4}.
We do not include the topologies already encountered in the computation
of the electro-weak form factor \cite{UgoRo1,UgoRo2}.
By shrinking a second internal line, we obtain the four denominator
topologies shown in Fig.~\ref{fig5}.
There are no new three denominator topologies. 
By using the integration-by-parts identities, the topologies shown in
Figs.~\ref{fig3}--\ref{fig5} are reduced to the eight MIs
shown in Fig.~\ref{fig7}. 
All the irreducible topologies have only one MI, with the exception of 
the topology (a) in Fig.~\ref{fig5}, which has two MIs. 
As is often the case \cite{UgoRo1}, the non-planar topology is the only 
irreducible one among the six-denominator amplitudes.

All the MIs are computed with the differential equation technique
introduced in \cite{Kotikov1} and applied to similar cases
as the present one in \cite{UgoRo1,UgoRo2}. For the topology (a) in 
Fig.~\ref{fig5}, we choose the MIs consisting of the scalar 
amplitude and the amplitude with a massless denominator 
squared, given in (g) and (h) of Fig.~\ref{fig7} respectively.
With this choice, the system of two differential equations 
is triangular in four dimensions, allowing for an elementary
solution. 

The MIs that we present, are regularized within
the dimensional regularization scheme \cite{DimReg}.
They are expanded in a Laurent series of
$\epsilon\, = \, 2-D/2$, where $D$ is the space-time dimension.
We work in Minkowski space and the loop measure is normalized as:
${\mathfrak D}^D k = d^D k / \left[ i\pi^{D/2} 
\Gamma\left( 3 - D/2 \right) \right]$. We have defined 
$x\, = \, -s/a$, where $a=m_{W,\, Z}^2$ and $s = -(p_1+p_2)^2$ 
\footnote{Scalar products are defined as: 
$a\cdot b = -a_0 b_0+\vec{a}\cdot\vec{b}$.} 
is the c.m. energy squared, with $p_1$ and $p_2$ the light-cone momenta
of the final photons. The scale $\mu$ is, as usual, the unit of mass of 
dimensional regularization. The results are naturally expressed as a 
linear combination of generalized harmonic polylogarithms (GHPLs) 
\cite{Polylog,Polylog3,UgoRo2} of the variable $x$, with maximum 
weight four included. The definition of the generalized harmonic 
polylogarithms and all the relevant conventions have been given 
in \cite{UgoRo1,UgoRo2}, to which we refer for details.

In the following we list the MIs in order of increasing number 
$t$ of denominators. We provide also a small appendix with the
expressions of the one-loop results entering the renormalization
of the two-loop corrections to $H \rightarrow \gamma \gamma$.

\bfig
\bc
\subfigure[]{
\begin{fmfgraph*}(25,25)
\fmfleft{i1,i2}
\fmfright{o}
\fmfforce{0.2w,0.08h}{v1}
\fmfforce{0.2w,0.92h}{v2}
\fmfforce{0.82w,0.5h}{v5}
\fmf{gluon}{i1,v1}
\fmf{gluon}{i2,v2}
\fmf{dashes}{v5,o}
\fmflabel{$g$}{i1}
\fmflabel{$g$}{i2}
\fmflabel{$H$}{o}
\fmf{photon,tension=.3,label=$W$,
                       label.side=left}{v3,v5}
\fmf{photon,tension=.3,label=$W$,
                       label.side=right}{v4,v5}
\fmf{fermion,tension=.3,label=$f$,
                       label.side=right}{v3,v2}
\fmf{fermion,tension=.3,label=$f$,
                       label.side=right}{v1,v4}
\fmf{fermion,tension=0,label=$f$,label.side=right}{v2,v1}
\fmf{fermion,tension=0,label=$f'$,label.side=left}{v4,v3}
\end{fmfgraph*} }
%
%
\hspace{6mm}
\subfigure[]{
\begin{fmfgraph*}(25,25)
\fmfleft{i1,i2}
\fmfright{o}
\fmfforce{0.2w,0.08h}{v1}
\fmfforce{0.2w,0.92h}{v2}
\fmfforce{0.82w,0.5h}{v5}
\fmf{gluon}{i1,v1}
\fmf{gluon}{i2,v2}
\fmf{dashes}{v5,o}
\fmflabel{$g$}{i1}
\fmflabel{$g$}{i2}
\fmflabel{$H$}{o}
\fmf{photon,tension=.3,label=$Z^{\circ}$,
                       label.side=left}{v3,v5}
\fmf{photon,tension=.3,label=$Z^{\circ}$,
                       label.side=right}{v4,v5}
\fmf{fermion,tension=.3,label=$f$,
                       label.side=right}{v3,v2}
\fmf{fermion,tension=.3,label=$f$,
                       label.side=right}{v1,v4}
\fmf{fermion,tension=0,label=$f$,label.side=right}{v2,v1}
\fmf{fermion,tension=0,label=$f$,label.side=left}{v4,v3}
\end{fmfgraph*} }
%
%
\hspace{6mm}
\subfigure[]{
\begin{fmfgraph*}(25,25)
\fmfleft{i1,i2}
\fmfright{o}
\fmfforce{0.2w,0.08h}{v1}
\fmfforce{0.2w,0.92h}{v2}
\fmfforce{0.82w,0.5h}{v5}
\fmfforce{0.3w,0.6h}{v7}
\fmfforce{0.3w,0.4h}{v8}
\fmf{gluon}{i1,v1}
\fmf{gluon}{i2,v2}
\fmf{dashes}{v5,o}
\fmflabel{$g$}{i1}
\fmflabel{$g$}{i2}
\fmflabel{$H$}{o}
\fmfv{label=$f$,label.side=left}{v7}
\fmfv{label=$f'$,label.side=left}{v8}
\fmf{photon,tension=.3,label=$W$,
                       label.side=left}{v3,v5}
\fmf{photon,tension=.3,label=$W$,
                       label.side=right}{v4,v5}
\fmf{fermion,tension=.3,label=$f$,
                       label.side=right}{v3,v2}
\fmf{fermion,tension=.3,label=$f'$,
                       label.side=left}{v4,v1}
\fmf{fermion,tension=0}{v2,v4}
\fmf{fermion,tension=0}{v1,v3}
\end{fmfgraph*} }
%
%
\hspace{6mm}
\subfigure[]{
\begin{fmfgraph*}(25,25)
\fmfleft{i1,i2}
\fmfright{o}
\fmfforce{0.2w,0.08h}{v1}
\fmfforce{0.2w,0.92h}{v2}
\fmfforce{0.82w,0.5h}{v5}
\fmfforce{0.3w,0.6h}{v7}
\fmfforce{0.3w,0.4h}{v8}
\fmf{gluon}{i1,v1}
\fmf{gluon}{i2,v2}
\fmf{dashes}{v5,o}
\fmflabel{$g$}{i1}
\fmflabel{$g$}{i2}
\fmflabel{$H$}{o}
\fmfv{label=$f$,label.side=left}{v7}
\fmfv{label=$f$,label.side=left}{v8}
\fmf{photon,tension=.3,label=$Z^{\circ}$,
                       label.side=left}{v3,v5}
\fmf{photon,tension=.3,label=$Z^{\circ}$,
                       label.side=right}{v4,v5}
\fmf{fermion,tension=.3,label=$f$,
                       label.side=right}{v3,v2}
\fmf{fermion,tension=.3,label=$f$,
                       label.side=left}{v4,v1}
\fmf{fermion,tension=0}{v2,v4}
\fmf{fermion,tension=0}{v1,v3}
\end{fmfgraph*} } 
%
%
\vspace*{4mm}
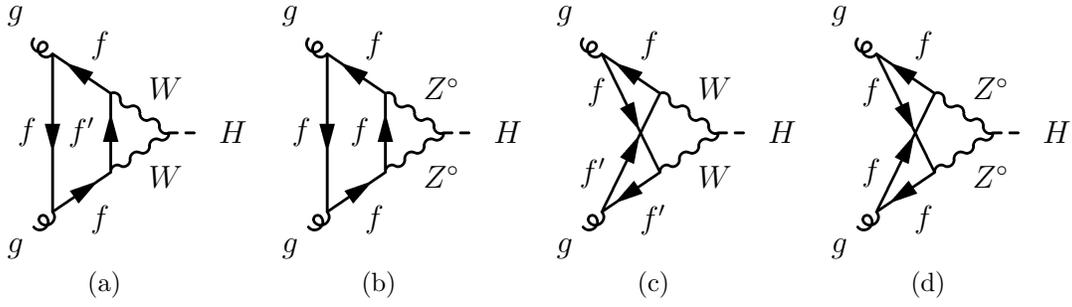
\caption{\label{fig1} Feynman diagrams for electro-weak light fermion
contributions to $gg \rightarrow H$. The straight lines represent light 
fermions, while the wavy lines stand for the $W$ or $Z$ bosons. The 
curly lines denote the initial gluons, the dashed line the final Higgs 
boson.}
\ec
\efig

\bfig
\bc
\subfigure[]{
\begin{fmfgraph*}(25,25)
\fmfleft{i}
\fmfright{o1,o2}
\fmfforce{0.14w,0.5h}{v1}
\fmf{dashes}{i,v1}
\fmf{photon}{v4,o1}
\fmf{photon}{v5,o2}
\fmflabel{$H$}{i}
\fmflabel{$\gamma$}{o1}
\fmflabel{$\gamma$}{o2}
\fmf{photon,tension=.3,label=$Z^{\circ}$,label.side=right}{v1,v2}
\fmf{photon,tension=.3,label=$Z^{\circ}$,label.side=left}{v1,v3}
\fmf{fermion,tension=.3,label=$f$,label.side=left}{v4,v2}
\fmf{fermion,tension=.3,label=$f$,label.side=left}{v3,v5}
\fmf{fermion,tension=0,label=$f$,label.side=right}{v2,v3}
\fmf{fermion,tension=0,label=$f$,label.side=left}{v5,v4}
\end{fmfgraph*} }
%
%
\hspace{6mm}
\subfigure[]{
\begin{fmfgraph*}(25,25)
\fmfleft{i}
\fmfright{o1,o2}
\fmfforce{0.14w,0.5h}{v1}
\fmf{dashes}{i,v1}
\fmf{photon}{v4,o1}
\fmf{photon}{v5,o2}
\fmflabel{$H$}{i}
\fmflabel{$\gamma$}{o1}
\fmflabel{$\gamma$}{o2}
\fmf{photon,tension=.3,label=$W$,label.side=right}{v1,v2}
\fmf{photon,tension=.3,label=$W$,label.side=left}{v1,v3}
\fmf{fermion,tension=.3,label=$f$,label.side=left}{v4,v2}
\fmf{fermion,tension=.3,label=$f$,label.side=left}{v3,v5}
\fmf{fermion,tension=0,label=$f'$,label.side=right}{v2,v3}
\fmf{fermion,tension=0,label=$f$,label.side=left}{v5,v4}
\end{fmfgraph*} }
%
%
\hspace{6mm}
\subfigure[]{
\begin{fmfgraph*}(25,25)
\fmfleft{i}
\fmfright{o1,o2}
\fmfforce{0.14w,0.5h}{v1}
\fmfforce{0.68w,0.6h}{v7}
\fmfforce{0.68w,0.4h}{v8}
\fmf{dashes}{i,v1}
\fmf{photon}{v4,o1}
\fmf{photon}{v5,o2}
\fmflabel{$H$}{i}
\fmflabel{$\gamma$}{o1}
\fmflabel{$\gamma$}{o2}
\fmf{photon,tension=.3,label=$Z^{\circ}$,label.side=right}{v1,v2}
\fmf{photon,tension=.3,label=$Z^{\circ}$,label.side=left}{v1,v3}
\fmf{fermion,tension=.3,label=$f$,label.side=left}{v4,v2}
\fmf{fermion,tension=.3,label=$f$,label.side=right}{v5,v3}
\fmf{fermion,tension=0}{v2,v5}
\fmf{fermion,tension=0}{v3,v4}
\fmfv{label=$f$,label.side=right}{v7}
\fmfv{label=$f$,label.side=right}{v8}
\end{fmfgraph*} }
%
%
\hspace{6mm} 
\subfigure[]{
\begin{fmfgraph*}(25,25)
\fmfleft{i}
\fmfright{o1,o2}
\fmfforce{0.14w,0.5h}{v1}
\fmfforce{0.68w,0.6h}{v7}
\fmfforce{0.68w,0.4h}{v8}
\fmf{dashes}{i,v1}
\fmf{photon}{v4,o1}
\fmf{photon}{v5,o2}
\fmflabel{$H$}{i}
\fmflabel{$\gamma$}{o1}
\fmflabel{$\gamma$}{o2}
\fmf{photon,tension=.3,label=$W$,label.side=right}{v1,v2}
\fmf{photon,tension=.3,label=$W$,label.side=left}{v1,v3}
\fmf{fermion,tension=.3,label=$f'$,label.side=left}{v4,v2}
\fmf{fermion,tension=.3,label=$f$,label.side=right}{v5,v3}
\fmf{fermion,tension=0}{v2,v5}
\fmf{fermion,tension=0}{v3,v4}
\fmfv{label=$f$,label.side=right}{v7}
\fmfv{label=$f'$,label.side=right}{v8}
\end{fmfgraph*} } \\
%
%
\subfigure[]{
\begin{fmfgraph*}(25,25)
\fmfleft{i}
\fmfright{o1,o2}
\fmfforce{0.14w,0.5h}{v1}
\fmfforce{0.8w,0.5h}{v3}
\fmfforce{0.8w,0.9h}{v5}
\fmfforce{0.8w,0.1h}{v4}
\fmf{dashes}{i,v1}
\fmf{photon}{v4,o1}
\fmf{photon}{v5,o2}
\fmflabel{$H$}{i}
\fmflabel{$\gamma$}{o1}
\fmflabel{$\gamma$}{o2}
\fmf{photon,tension=0,label=$W$,label.side=left}{v1,v5}
\fmf{photon,tension=.4,label=$W$,label.side=right}{v1,v2}
\fmf{fermion,tension=.4,label=$f$,label.side=left}{v4,v2}
\fmf{fermion,tension=0,label=$f'$,label.side=left}{v2,v3}
\fmf{fermion,tension=0,label=$f$,label.side=left}{v3,v4}
\fmf{photon,tension=0,label=$W$,label.side=right}{v3,v5}
\end{fmfgraph*} }
%
%
\hspace{6mm}
\subfigure[]{
\begin{fmfgraph*}(25,25)
\fmfleft{i}
\fmfright{o1,o2}
\fmfforce{0.14w,0.5h}{v1}
\fmfforce{0.8w,0.5h}{v3}
\fmfforce{0.8w,0.9h}{v5}
\fmfforce{0.8w,0.1h}{v4}
\fmf{dashes}{i,v1}
\fmf{photon}{v4,o1}
\fmf{photon}{v5,o2}
\fmflabel{$H$}{i}
\fmflabel{$\gamma$}{o1}
\fmflabel{$\gamma$}{o2}
\fmf{photon,tension=0,label=$W$,label.side=right}{v1,v4}
\fmf{photon,tension=.4,label=$W$,label.side=left}{v1,v2}
\fmf{fermion,tension=.4,label=$f$,label.side=right}{v5,v2}
\fmf{fermion,tension=0,label=$f'$,label.side=right}{v2,v3}
\fmf{fermion,tension=0,label=$f$,label.side=right}{v3,v5}
\fmf{photon,tension=0,label=$W$,label.side=left}{v3,v4}
\end{fmfgraph*} }
%
%
\hspace{6mm}
\subfigure[]{
\begin{fmfgraph*}(25,25)
\fmfleft{i}
\fmfright{o1,o2}
\fmfforce{0.14w,0.5h}{v1}
\fmfforce{0.8w,0.9h}{v5}
\fmfforce{0.8w,0.1h}{v4}
\fmf{dashes}{i,v1}
\fmf{photon}{v4,o1}
\fmf{photon}{v5,o2}
\fmflabel{$H$}{i}
\fmflabel{$\gamma$}{o1}
\fmflabel{$\gamma$}{o2}
\fmf{photon,tension=.3,label=$W$,label.side=left}{v1,v3}
\fmf{fermion,tension=.15,left,label=$f'$,label.side=left}{v3,v6}
\fmf{fermion,tension=.15,left,label=$f$,label.side=left}{v6,v3}
\fmf{photon,tension=.3,label=$W$,label.side=left}{v6,v5}
\fmf{photon,tension=.3,label=$W$,label.side=right}{v1,v4}
\fmf{photon,tension=.3,label=$W$,label.side=left}{v5,v4}
\fmf{photon,tension=0}{v5,v4}
\end{fmfgraph*} }
%
%
\hspace{6mm}
\subfigure[]{
\begin{fmfgraph*}(25,25)
\fmfleft{i}
\fmfright{o1,o2}
\fmfforce{0.14w,0.5h}{v1}
\fmfforce{0.8w,0.9h}{v5}
\fmfforce{0.8w,0.1h}{v4}
\fmf{dashes}{i,v1}
\fmf{photon}{v4,o1}
\fmf{photon}{v5,o2}
\fmflabel{$H$}{i}
\fmflabel{$\gamma$}{o1}
\fmflabel{$\gamma$}{o2}
\fmf{photon,tension=.3,label=$W$,label.side=right}{v1,v3}
\fmf{fermion,tension=.15,left,label=$f$,label.side=left}{v3,v6}
\fmf{fermion,tension=.15,left,label=$f'$,label.side=left}{v6,v3}
\fmf{photon,tension=.3,label=$W$,label.side=right}{v6,v4}
\fmf{photon,tension=.3,label=$W$,label.side=left}{v1,v5}
\fmf{photon,tension=.3,label=$W$,label.side=left}{v5,v4}
\end{fmfgraph*} } \\
%
%
\subfigure[]{
\begin{fmfgraph*}(25,25)
\fmfleft{i}
\fmfright{o1,o2}
\fmfforce{0.14w,0.5h}{v1}
\fmfforce{0.8w,0.9h}{v5}
\fmfforce{0.8w,0.1h}{v4}
\fmf{dashes}{i,v1}
\fmf{photon}{v4,o1}
\fmf{photon}{v5,o2}
\fmflabel{$H$}{i}
\fmflabel{$\gamma$}{o1}
\fmflabel{$\gamma$}{o2}
\fmf{photon,tension=.3,label=$W$,label.side=right}{v1,v4}
\fmf{photon,tension=.3,label=$W$,label.side=left}{v5,v3}
\fmf{fermion,tension=.15,left,label=$f'$,label.side=left}{v3,v6}
\fmf{fermion,tension=.15,left,label=$f$,label.side=left}{v6,v3}
\fmf{photon,tension=.3,label=$W$,label.side=left}{v6,v4}
\fmf{photon,tension=.3,label=$W$,label.side=left}{v1,v5}
\end{fmfgraph*} }
%
%
\hspace{6mm}
\subfigure[]{
\begin{fmfgraph*}(25,25)
\fmfleft{i}
\fmfright{o1,o2}
\fmfforce{0.14w,0.5h}{v1}
\fmfforce{0.25w,0.74h}{v2}
\fmfforce{0.55w,0.74h}{v6}
\fmfforce{0.7w,0.5h}{v4}
\fmf{dashes}{i,v1}
\fmf{photon}{v4,o1}
\fmf{photon}{v4,o2}
\fmflabel{$H$}{i}
\fmflabel{$\gamma$}{o1}
\fmflabel{$\gamma$}{o2}
\fmf{photon,left=.3,label=$W$,label.side=left}{v1,v2}
\fmf{fermion,tension=.25,left,label=$f'$,label.side=left}{v2,v6}
\fmf{fermion,tension=.25,left,label=$f$,label.side=left}{v6,v2}
\fmf{photon,left=.3}{v6,v4}
\fmf{photon,right,label=$W$,label.side=right}{v1,v4}
\end{fmfgraph*} }
%
%
\hspace{6mm}
\subfigure[]{
\begin{fmfgraph*}(25,25)
\fmfleft{i}
\fmfright{o1,o2}
\fmfforce{0.14w,0.5h}{v1}
\fmfforce{0.25w,0.26h}{v2}
\fmfforce{0.55w,0.26h}{v6}
\fmfforce{0.7w,0.5h}{v4}
\fmf{dashes}{i,v1}
\fmf{photon}{v4,o1}
\fmf{photon}{v4,o2}
\fmflabel{$H$}{i}
\fmflabel{$\gamma$}{o1}
\fmflabel{$\gamma$}{o2}
\fmf{photon,right=.3,label=$W$,label.side=right}{v1,v2}
\fmf{fermion,tension=.25,left,label=$f$,label.side=left}{v2,v6}
\fmf{fermion,tension=.25,left,label=$f'$,label.side=left}{v6,v2}
\fmf{photon,right=.3}{v6,v4}
\fmf{photon,left,label=$W$,label.side=left}{v1,v4}
\end{fmfgraph*} }
%
%
\vspace*{4mm}
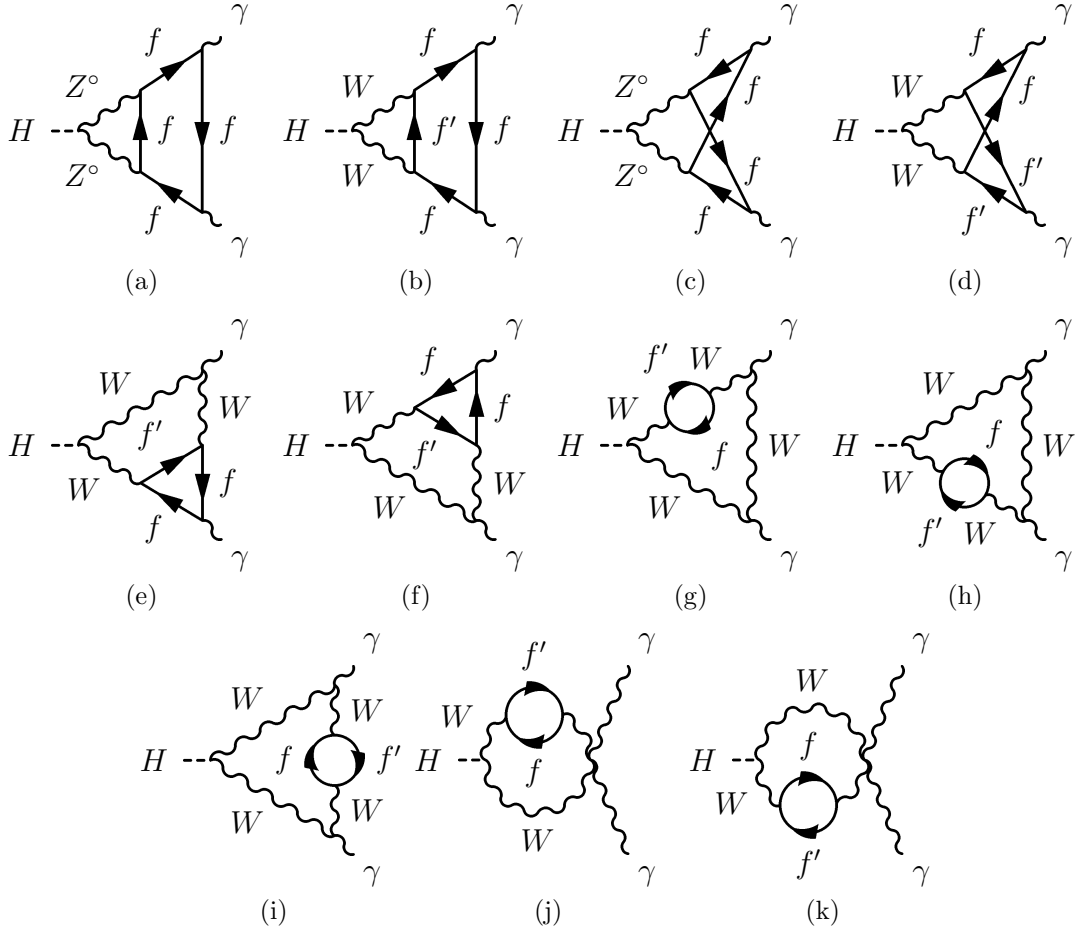
\caption{\label{fig2} Feynman diagrams for electro-weak 
light fermion contributions to the decay $H \rightarrow \gamma \gamma$.}
\ec
\efig

\bfig
\bc
\subfigure[]{
\begin{fmfgraph*}(20,20)
\fmfleft{i}
\fmfright{o1,o2}
\fmfforce{0.1w,0.5h}{v1}
\fmf{dashes}{i,v1}
\fmf{photon}{v4,o1}
\fmf{photon}{v5,o2}
\fmf{plain,tension=.3}{v1,v2}
\fmf{plain,tension=.3}{v1,v3}
\fmf{photon,tension=.3}{v2,v4}
\fmf{photon,tension=.3}{v3,v5}
\fmf{photon,tension=0}{v2,v3}
\fmf{photon,tension=0}{v4,v5}
\end{fmfgraph*} }
%
%
\hspace{6mm}
\subfigure[]{
\begin{fmfgraph*}(20,20)
\fmfleft{i}
\fmfright{o1,o2}
\fmfforce{0.1w,0.5h}{v1}
\fmf{dashes}{i,v1}
\fmf{photon}{v4,o1}
\fmf{photon}{v5,o2}
\fmf{plain,tension=.3}{v1,v2}
\fmf{plain,tension=.3}{v1,v3}
\fmf{photon,tension=.3}{v2,v4}
\fmf{photon,tension=.3}{v3,v5}
\fmf{photon,tension=0}{v2,v5}
\fmf{photon,tension=0}{v3,v4}
\end{fmfgraph*} }
%
%
\hspace{6mm}
\subfigure[]{
\begin{fmfgraph*}(20,20)
\fmfleft{i}
\fmfright{o1,o2}
\fmfforce{0.1w,0.5h}{v1}
\fmfforce{0.8w,0.5h}{v3}
\fmfforce{0.8w,0.9h}{v5}
\fmfforce{0.8w,0.1h}{v4}
\fmf{dashes}{i,v1}
\fmf{photon}{v4,o1}
\fmf{photon}{v5,o2}
\fmf{plain,tension=0}{v1,v5}
\fmf{plain,tension=.4}{v1,v2}
\fmf{photon,tension=.4}{v2,v4}
\fmf{photon,tension=0}{v2,v3}
\fmf{photon,tension=0}{v4,v3}
\fmf{plain,tension=0}{v3,v5}
\end{fmfgraph*} } 
%
\vspace*{4mm}
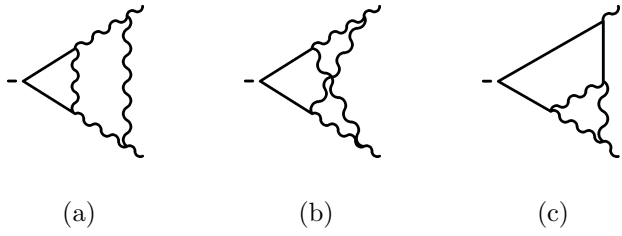
\caption{\label{fig3} The set of three independent topologies with six
denominators, related to the Feynman diagrams shown in Figs.~\ref{fig1} 
and \ref{fig2}.}
\ec
\efig

\bfig
\bc
\subfigure[]{
\begin{fmfgraph*}(20,20)
\fmfleft{i}
\fmfright{o1,o2}
\fmfforce{0.2w,0.5h}{v1}
\fmf{dashes}{i,v1}
\fmf{photon}{v4,o1}
\fmf{photon}{v5,o2}
\fmf{plain,tension=.2}{v1,v3}
\fmf{photon,tension=.4}{v3,v5}
\fmf{plain,tension=.15}{v1,v4}
\fmf{photon,tension=0}{v4,v5}
\fmf{photon,tension=0}{v4,v3}
\end{fmfgraph*} }  
%
%
\hspace{5mm}
\subfigure[]{
\begin{fmfgraph*}(20,20)
\fmfleft{i}
\fmfright{o1,o2}
\fmfforce{0.2w,0.5h}{v1}
\fmf{dashes}{i,v1}
\fmf{photon}{v4,o1}
\fmf{photon}{v5,o2}
\fmf{photon,tension=.2}{v1,v2}
\fmf{photon,tension=0,left=.5}{v1,v2}
\fmf{plain,tension=.15}{v1,v5}
\fmf{plain,tension=.4}{v2,v4}
\fmf{plain,tension=0}{v4,v5}
\end{fmfgraph*} }
%
%
\hspace{5mm}
\subfigure[]{
\begin{fmfgraph*}(20,20)
\fmfleft{i}
\fmfright{o1,o2}
\fmfforce{0.2w,0.5h}{v1}
\fmfforce{0.83w,0.5h}{v3}
\fmfforce{0.83w,0.13h}{v4}
\fmfforce{0.83w,0.87h}{v5}
\fmf{dashes}{i,v1}
\fmf{photon}{v4,o1}
\fmf{photon}{v5,o2}
\fmf{plain,tension=0}{v1,v3}
\fmf{photon,tension=0}{v1,v4}
\fmf{photon,tension=0}{v1,v5}
\fmf{photon,tension=0}{v4,v3}
\fmf{photon,tension=0}{v5,v3}
\end{fmfgraph*} }
%
%
\hspace{5mm}
\subfigure[]{
\begin{fmfgraph*}(20,20)
\fmfleft{i}
\fmfright{o1,o2}
\fmfforce{0.2w,0.5h}{v1}
\fmfforce{0.83w,0.5h}{v3}
\fmfforce{0.83w,0.13h}{v4}
\fmfforce{0.83w,0.87h}{v5}
\fmfforce{0.83w,0.17h}{v6}
\fmf{dashes}{i,v1}
\fmf{photon}{v4,o1}
\fmf{photon}{v5,o2}
\fmf{plain}{v1,v4}
\fmf{plain}{v1,v5}
\fmf{photon,left}{v3,v6}
\fmf{photon,right}{v3,v6}
\fmf{plain}{v5,v3}
\end{fmfgraph*} } \\
%
%
\subfigure[]{
\begin{fmfgraph*}(20,20)
\fmfleft{i}
\fmfright{o1,o2}
\fmfforce{0.2w,0.5h}{v1}
\fmfforce{0.83w,0.5h}{v3}
\fmfforce{0.83w,0.13h}{v4}
\fmfforce{0.83w,0.87h}{v5}
\fmf{dashes}{i,v1}
\fmf{photon}{v4,o1}
\fmf{photon}{v5,o2}
\fmf{photon,tension=0}{v1,v3}
\fmf{photon,tension=0}{v1,v4}
\fmf{plain,tension=0}{v1,v5}
\fmf{photon,tension=0}{v4,v3}
\fmf{plain,tension=0}{v5,v3}
\end{fmfgraph*} }
%
%
\hspace{9mm}
\subfigure[]{
\begin{fmfgraph*}(20,20)
\fmfforce{0.2w,0.5h}{v1}
\fmfforce{0.5w,0.8h}{v2}
\fmfforce{0.5w,0.2h}{v3}
\fmfforce{0.8w,0.5h}{v4}
\fmfleft{i}
\fmfright{o}
\fmf{photon}{i,v1}
\fmf{photon}{v4,o}
\fmf{photon,tension=.2,left=.4}{v1,v2}
\fmf{photon,tension=.2,right=.4}{v1,v3}
\fmf{plain,tension=.2,left=.4}{v2,v4}
\fmf{plain,tension=.2,right=.4}{v3,v4}
\fmf{photon,tension=0}{v2,v3}
\end{fmfgraph*} }
%
%
\vspace*{4mm}
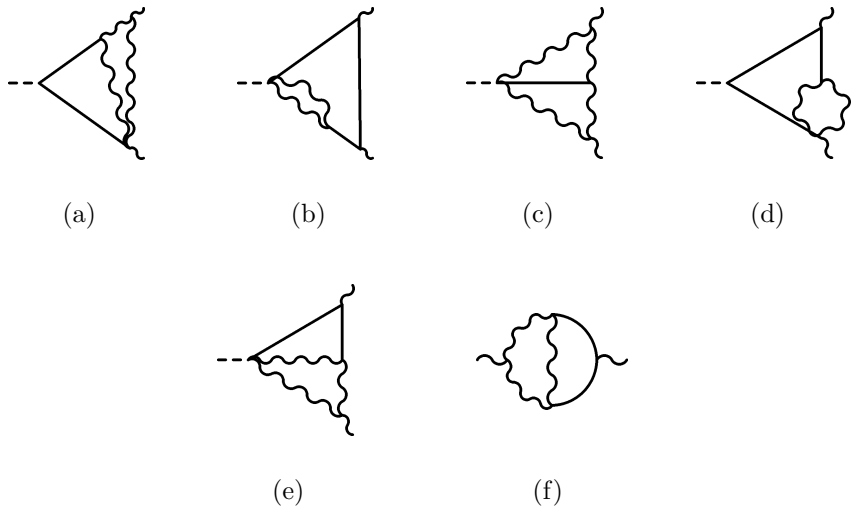
\caption{\label{fig4} The set of six independent 
five-denominator topologies.}
\ec
\efig

\bfig
\bc
%
\subfigure[]{
\begin{fmfgraph*}(20,20)
\fmfleft{i}
\fmfright{o1,o2}
\fmfforce{0.2w,0.5h}{v1}
\fmf{dashes}{i,v1}
\fmf{photon}{v2,o1}
\fmf{photon}{v3,o2}
\fmf{plain,tension=.3}{v1,v3}
\fmf{plain,tension=.3}{v1,v2}
\fmf{photon,tension=0,right=.5}{v2,v3}
\fmf{photon,tension=0,right=.5}{v3,v2}
\end{fmfgraph*} }
%
%
\hspace{9mm}
\subfigure[]{
\begin{fmfgraph*}(20,20)
\fmfleft{i}
\fmfright{o1,o2}
\fmfforce{0.2w,0.5h}{v1}
\fmf{dashes}{i,v1}
\fmf{photon}{v2,o1}
\fmf{photon}{v3,o2}
\fmf{photon,tension=.3}{v1,v2}
\fmf{photon,tension=0,right=.5}{v2,v1}
\fmf{plain,tension=.3}{v1,v3}
\fmf{plain,tension=0}{v2,v3}
\end{fmfgraph*} } 
%
%
\hspace{9mm}
\subfigure[]{
\begin{fmfgraph*}(20,20)
\fmfleft{i}
\fmfright{o}
\fmfforce{0.2w,0.5h}{v1}
\fmfforce{0.5w,0.2h}{v2}
\fmfforce{0.8w,0.5h}{v3}
\fmf{photon}{i,v1}
\fmf{photon}{v3,o}
\fmf{plain,left}{v1,v3}
\fmf{plain,right=.4}{v1,v2}
\fmf{photon,right=.4}{v2,v3}
\fmf{photon,left=.6}{v2,v3}
\end{fmfgraph*} } 
%
%
\vspace*{4mm}
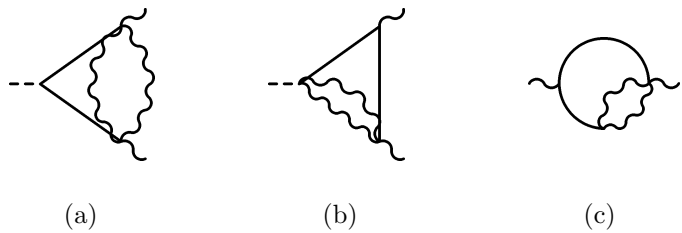
\caption{\label{fig5} The set of three independent 
four-denominator topologies.}
\ec
\efig

\bfig
\bc
\subfigure[]{
\begin{fmfgraph*}(20,20)
\fmfleft{i}
\fmfright{o1,o2}
\fmfforce{0.2w,0.5h}{v1}
\fmf{dashes}{i,v1}
\fmf{photon}{v4,o1}
\fmf{photon}{v5,o2}
\fmf{plain,tension=.3}{v1,v2}
\fmf{plain,tension=.3}{v1,v3}
\fmf{photon,tension=.3}{v2,v4}
\fmf{photon,tension=.3}{v3,v5}
\fmf{photon,tension=0}{v2,v5}
\fmf{photon,tension=0}{v3,v4}
\end{fmfgraph*} } 
%
%
\hspace{5mm}
\subfigure[]{
\begin{fmfgraph*}(20,20)
\fmfleft{i}
\fmfright{o1,o2}
\fmfforce{0.2w,0.5h}{v1}
\fmf{dashes}{i,v1}
\fmf{photon}{v4,o1}
\fmf{photon}{v5,o2}
\fmf{plain,tension=.2}{v1,v3}
\fmf{photon,tension=.4}{v3,v5}
\fmf{plain,tension=.15}{v1,v4}
\fmf{photon,tension=0}{v4,v5}
\fmf{photon,tension=0}{v4,v3}
\end{fmfgraph*} }  
%
%
\hspace{5mm}
\subfigure[]{
\begin{fmfgraph*}(20,20)
\fmfleft{i}
\fmfright{o1,o2}
\fmfforce{0.2w,0.5h}{v1}
\fmf{dashes}{i,v1}
\fmf{photon}{v4,o1}
\fmf{photon}{v5,o2}
\fmf{photon,tension=.2}{v1,v2}
\fmf{photon,tension=0,left=.5}{v1,v2}
\fmf{plain,tension=.15}{v1,v5}
\fmf{plain,tension=.4}{v2,v4}
\fmf{plain,tension=0}{v4,v5}
\end{fmfgraph*} }
%
%
\hspace{5mm}
\subfigure[]{
\begin{fmfgraph*}(20,20)
\fmfleft{i}
\fmfright{o1,o2}
\fmfforce{0.2w,0.5h}{v1}
\fmfforce{0.83w,0.5h}{v3}
\fmfforce{0.83w,0.13h}{v4}
\fmfforce{0.83w,0.87h}{v5}
\fmf{dashes}{i,v1}
\fmf{photon}{v4,o1}
\fmf{photon}{v5,o2}
\fmf{plain,tension=0}{v1,v3}
\fmf{photon,tension=0}{v1,v4}
\fmf{photon,tension=0}{v1,v5}
\fmf{photon,tension=0}{v4,v3}
\fmf{photon,tension=0}{v5,v3}
\end{fmfgraph*} } \\
%
%
\subfigure[]{
\begin{fmfgraph*}(20,20)
\fmfleft{i}
\fmfright{o1,o2}
\fmfforce{0.2w,0.5h}{v1}
\fmfforce{0.83w,0.5h}{v3}
\fmfforce{0.83w,0.13h}{v4}
\fmfforce{0.83w,0.87h}{v5}
\fmfforce{0.83w,0.17h}{v6}
\fmf{dashes}{i,v1}
\fmf{photon}{v4,o1}
\fmf{photon}{v5,o2}
\fmf{plain}{v1,v4}
\fmf{plain}{v1,v5}
\fmf{photon,left}{v3,v6}
\fmf{photon,right}{v3,v6}
\fmf{plain}{v5,v3}
\end{fmfgraph*} }
%
%
\hspace{5mm}
\subfigure[]{
\begin{fmfgraph*}(20,20)
\fmfleft{i}
\fmfright{o1,o2}
\fmfforce{0.2w,0.5h}{v1}
\fmfforce{0.83w,0.5h}{v3}
\fmfforce{0.83w,0.13h}{v4}
\fmfforce{0.83w,0.87h}{v5}
\fmf{dashes}{i,v1}
\fmf{photon}{v4,o1}
\fmf{photon}{v5,o2}
\fmf{photon,tension=0}{v1,v3}
\fmf{photon,tension=0}{v1,v4}
\fmf{plain,tension=0}{v1,v5}
\fmf{photon,tension=0}{v4,v3}
\fmf{plain,tension=0}{v5,v3}
\end{fmfgraph*} } 
%
%
\hspace{5mm}
\subfigure[]{
\begin{fmfgraph*}(20,20)
\fmfleft{i}
\fmfright{o1,o2}
\fmfforce{0.2w,0.5h}{v1}
\fmf{dashes}{i,v1}
\fmf{photon}{v2,o1}
\fmf{photon}{v3,o2}
\fmf{plain,tension=.3}{v1,v3}
\fmf{plain,tension=.3}{v1,v2}
\fmf{photon,tension=0,right=.5}{v2,v3}
\fmf{photon,tension=0,right=.5}{v3,v2}
\end{fmfgraph*} }
%
%
\hspace{5mm}
\subfigure[]{
\begin{fmfgraph*}(20,20)
\fmfleft{i}
\fmfright{o1,o2}
\fmfforce{0.2w,0.5h}{v1}
\fmfforce{0.93w,0.5h}{v10}
\fmf{dashes}{i,v1}
\fmf{photon}{v2,o1}
\fmf{photon}{v3,o2}
\fmf{plain,tension=.3}{v1,v3}
\fmf{plain,tension=.3}{v1,v2}
\fmf{photon,tension=0,right=.5}{v2,v3}
\fmf{photon,tension=0,right=.5}{v3,v2}
\fmfv{decor.shape=circle,decor.filled=full,decor.size=.1w}{v10}
\end{fmfgraph*} }
%
%
\vspace*{4mm}
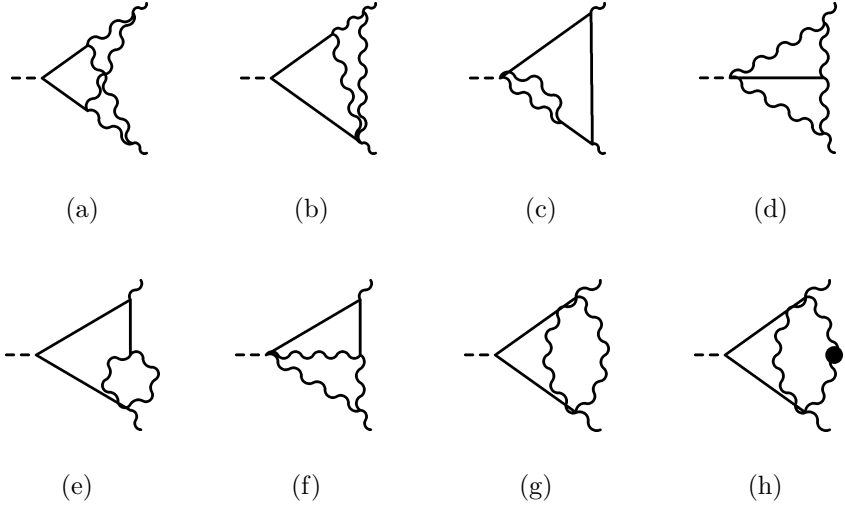
\caption{\label{fig7} The set of eight MIs. The dot on a line indicates 
a square of the corresponding denominator.}
\ec
\efig
%
%

\vskip 1.2truecm
\boldmath
\centerline{\bf Topology $t=4$ \label{4den}}
\unboldmath
\vskip 1truecm

\be
\parbox{20mm}{\begin{fmfgraph*}(15,15)
\fmfleft{i}
\fmfright{o1,o2}
\fmfforce{0.2w,0.5h}{v1}
\fmf{dashes}{i,v1}
\fmf{photon}{v2,o1}
\fmf{photon}{v3,o2}
\fmf{plain,tension=.3}{v1,v3}
\fmf{plain,tension=.3}{v1,v2}
\fmf{photon,tension=0,right=.5}{v2,v3}
\fmf{photon,tension=0,right=.5}{v3,v2}
\end{fmfgraph*}}  = \left( \frac{\mu^{2}}{a} \right) ^{2 \epsilon} 
\sum_{i=-2}^{2} \epsilon^{i} F^{(1)}_{i} + {\mathcal O} \left( 
\epsilon^{3} \right) , 
\ee
where:
\bea
F^{(1)}_{-2} & = & \frac{1}{2} \, , \\
F^{(1)}_{-1} & = & \frac{5}{2} 
                 - \frac{x+4}{\sqrt{x(x+4)}} H(-r;x)
 \, , \\
F^{(1)}_{0} & = & \frac{19}{2}  
                 + \zeta(2)
		 + H(-r,-r;x)
		 + \frac{2}{x} H(-r,-r;x)
                 - \frac{x+4}{\sqrt{x(x+4)}} \Bigl[ 
		   5 H(-r;x) \nn\\
& & 
		 - H(-4,-r;x) \Bigr]
\, , \\
F^{(1)}_{1} & = &  \frac{65}{2}
          + 5 \zeta(2)
          - \zeta(3)
          + 5 H( - r, - r,x)
          - H( - r,-4, - r,x) \nn\\
& & 
          + H(0, - r, - r,x)
       + \frac{1}{x} \Bigl[
            10 H( - r, - r,x)
          - 2 H( - r,-4, - r,x) \nn\\
& & 
          + 2 H(0, - r, - r,x)
          \Bigr]
       - \frac{x+4}{\sqrt{x(x+4)}} \Bigl[
            19 H( - r,x)
          + 2 \zeta(2) H( - r,x) \nn\\
& & 
          - 5 H(-4, - r,x)
          + 3 H( - r, - r, - r,x)
          + H(-4,-4, - r,x)
          \Bigr]
           \, , \\
F^{(1)}_{2} & = &  
            \frac{211}{2}
          + 19 \zeta(2)
          + \frac{9}{5} \zeta^2(2)
          - 5 \zeta(3)
          + 3 H( - r, - r, - r, - r,x)
          + 19 H( - r, - r,x)  \nn\\
& & 
          + 2 \zeta(2) H( - r, - r,x)
          - 5 H( - r,-4, - r,x)
          + H( - r,-4,-4, - r,x)  \nn\\
& & 
          + 5 H(0, - r, - r,x)
          - H(0, - r,-4, - r,x)
          + H(0,0, - r, - r,x)  \nn\\
& & 
       + \frac{1}{x} \Bigl[
            6 H( - r, - r, - r, - r,x)
          + 38 H( - r, - r,x)
          + 4 \zeta(2) H( - r, - r,x)  \nn\\
& & 
          - 10 H( - r,-4, - r,x)
          + 2 H( - r,-4,-4, - r,x)
          + 10 H(0, - r, - r,x)  \nn\\
& & 
          - 2 H(0, - r,-4, - r,x)
          + 2 H(0,0, - r, - r,x)
          \Bigr]  \nn\\
& & 
       - \frac{x+4}{\sqrt{x(x+4)}} \Bigl[
            65 H( - r,x)
          + 10 \zeta(2) H( - r,x)
          - 2 \zeta(3) H( - r,x)  \nn\\
& & 
          - 19 H(-4, - r,x)
          - 2 \zeta(2) H(-4, - r,x)
          + 15 H( - r, - r, - r,x)  \nn\\
& & 
          - 3 H( - r, - r,-4, - r,x)
          + 3 H( - r,0, - r, - r,x)
          - 3 H(-4, - r, - r, - r,x) \nn\\
& & 
          + 5 H(-4,-4, - r,x) 
          - H(-4,-4,-4, - r,x)
          \Bigr]
\, .
\eea

\be
\parbox{20mm}{\begin{fmfgraph*}(15,15)
\fmfleft{i}
\fmfright{o1,o2}
\fmfforce{0.2w,0.5h}{v1}
\fmfforce{0.93w,0.5h}{v10}
\fmf{dashes}{i,v1}
\fmf{photon}{v2,o1}
\fmf{photon}{v3,o2}
\fmf{plain,tension=.3}{v1,v3}
\fmf{plain,tension=.3}{v1,v2}
\fmf{photon,tension=0,right=.5}{v2,v3}
\fmf{photon,tension=0,right=.5}{v3,v2}
\fmfv{decor.shape=circle,decor.filled=full,decor.size=.1w}{v10}
\end{fmfgraph*}}  = \left( \frac{\mu^{2}}{a} \right) ^{2 \epsilon} 
\sum_{i=-1}^{1} \epsilon^{i} F^{(2)}_{i} + {\mathcal O} \left( 
\epsilon^{2} \right) , 
\ee
where:
\bea
aF^{(2)}_{-1} & = & - \frac{2}{x} H(-r,-r;x) \, , \\
aF^{(2)}_{0} & = & \frac{2}{x} \Bigl[ 
                  H(-r,-4,-r;x)
		- H(0,-r,-r;x) \Bigr] \, , \\
aF^{(2)}_{1} & = & - \frac{2}{x} \Bigl[  
                     2 \zeta(2) H(-r,-r;x)
		   + H(0,0,-r,-r;x)
		   + H(-r,-4,-4,-r;x) \nn\\
& & 
		   - H(0,-r,-4,-r;x)
		   + 3 H(-r,-r,-r,-r;x) \Bigr]
\, .
\eea

\vskip 1.3truecm
\boldmath
\centerline{\bf Topology $t=5$ \label{5den}}
\unboldmath
\vskip 0.8truecm

\be
\parbox{20mm}{\begin{fmfgraph*}(15,15)
\fmfleft{i}
\fmfright{o1,o2}
\fmfforce{0.2w,0.5h}{v1}
\fmfforce{0.83w,0.5h}{v3}
\fmfforce{0.83w,0.13h}{v4}
\fmfforce{0.83w,0.87h}{v5}
\fmf{dashes}{i,v1}
\fmf{photon}{v4,o1}
\fmf{photon}{v5,o2}
\fmf{photon,tension=0}{v1,v3}
\fmf{photon,tension=0}{v1,v4}
\fmf{plain,tension=0}{v1,v5}
\fmf{photon,tension=0}{v4,v3}
\fmf{plain,tension=0}{v5,v3}
\end{fmfgraph*}}  =  \left( \frac{\mu^{2}}{a} \right) ^{2 \epsilon} 
\sum_{i=-1}^{0} \epsilon^{i} F^{(3)}_{i} + {\mathcal O} \left( 
\epsilon \right) , 
\ee
where:
\bea
aF^{(3)}_{-1} & = &  - \frac{1}{x} H(0,0,-1,x)
\, , \\
aF^{(3)}_{0} & = &  \frac{1}{x} \Bigl[
            4 H(0,0,-1,-1,x)
          - H(0,0,0,-1,x) \Bigr]
 \, .
\eea

\be
\parbox{20mm}{\begin{fmfgraph*}(15,15)
\fmfleft{i}
\fmfright{o1,o2}
\fmfforce{0.2w,0.5h}{v1}
\fmfforce{0.83w,0.5h}{v3}
\fmfforce{0.83w,0.13h}{v4}
\fmfforce{0.83w,0.87h}{v5}
\fmfforce{0.83w,0.17h}{v6}
\fmf{dashes}{i,v1}
\fmf{photon}{v4,o1}
\fmf{photon}{v5,o2}
\fmf{plain}{v1,v4}
\fmf{plain}{v1,v5}
\fmf{photon,left}{v3,v6}
\fmf{photon,right}{v3,v6}
\fmf{plain}{v5,v3}
\end{fmfgraph*}}  =  \left( \frac{\mu^{2}}{a} \right) ^{2 \epsilon} 
\sum_{i=-1}^{1} \epsilon^{i} F^{(4)}_{i} + {\mathcal O} \left( 
\epsilon^{2} \right) , 
\ee
where:
\bea
aF^{(4)}_{-1} & = & \frac{1}{x} H( - r, - r,x)
\, , \\
aF^{(4)}_{0} & = &  \frac{1}{x} \Bigl[
                      2 H( - r, - r,x)
                    - H( - r,-4, - r,x) 
		    - H(0, - r, - r,x)
		     \Bigr]
\, , \\
aF^{(4)}_{1} & = & \frac{1}{x} \Bigl[
            2 ( \zeta(2) + 2 ) H( - r, - r,x)
          - 2 H( - r,-4, - r,x)
          - 2 H(0, - r, - r,x) \nn\\
& & 
          + 3 H( - r, - r, - r, - r,x)
          + H( - r,-4,-4, - r,x)
          + H(0, - r,-4, - r,x) \nn\\
& & 
          - H(0,0, - r, - r,x) \Bigr] \, .
\eea

\be
\parbox{20mm}{\begin{fmfgraph*}(15,15)
\fmfleft{i}
\fmfright{o1,o2}
\fmfforce{0.2w,0.5h}{v1}
\fmfforce{0.83w,0.5h}{v3}
\fmfforce{0.83w,0.13h}{v4}
\fmfforce{0.83w,0.87h}{v5}
\fmf{dashes}{i,v1}
\fmf{photon}{v4,o1}
\fmf{photon}{v5,o2}
\fmf{plain,tension=0}{v1,v3}
\fmf{photon,tension=0}{v1,v4}
\fmf{photon,tension=0}{v1,v5}
\fmf{photon,tension=0}{v4,v3}
\fmf{photon,tension=0}{v5,v3}
\end{fmfgraph*}}  =  \left( \frac{\mu^{2}}{a} \right) ^{2 \epsilon} 
\sum_{i=-2}^{0} \epsilon^{i} F^{(5)}_{i} + {\mathcal O} \left( 
\epsilon \right) , 
\ee
where:
\bea
aF^{(5)}_{-2} & = &  \frac{1}{x} H(0,-1,x)
\, , \\
aF^{(5)}_{-1} & = &  \frac{2}{x}  \Bigl[
            H(0,0,-1,x)
          - 2 H(0,-1,-1,x)
		     \Bigr]
\, , \\
aF^{(5)}_{0} & = &  \frac{1}{x}  \Bigl[
            2 \zeta(2) H(0,-1,x)
          + 16 H(0,-1,-1,-1,x)
          - 6 H(0,-1,0,-1,x) \nn\\
& & 
          - 8 H(0,0,-1,-1,x)
          + 2 H(0,0,0,-1,x)
		     \Bigr]
\, .
\eea

\be
\parbox{20mm}{\begin{fmfgraph*}(15,15)
\fmfleft{i}
\fmfright{o1,o2}
\fmfforce{0.2w,0.5h}{v1}
\fmf{dashes}{i,v1}
\fmf{photon}{v4,o1}
\fmf{photon}{v5,o2}
\fmf{photon,tension=.2}{v1,v2}
\fmf{photon,tension=0,left=.5}{v1,v2}
\fmf{plain,tension=.15}{v1,v5}
\fmf{plain,tension=.4}{v2,v4}
\fmf{plain,tension=0}{v4,v5}
\end{fmfgraph*}}  =  \left( \frac{\mu^{2}}{a} \right) ^{2 \epsilon} 
\sum_{i=-1}^{1} \epsilon^{i} F^{(6)}_{i} + {\mathcal O} \left( 
\epsilon^{2} \right) , 
\ee
where:
\bea
aF^{(6)}_{-1} & = & \frac{1}{x} H( - r, - r,x)
\, , \\
aF^{(6)}_{0} & = &  \frac{1}{x}  \Biggl[
            2 H( - r, - r,x)
          - \frac{3}{2} H( - r, - r,-1,x)
          - H( - r,-4, - r,x) \nn\\
& & 
          - \frac{1}{2} H(0,0,-1,x)
          \Biggr]
\, , \\
aF^{(6)}_{1} & = & \frac{1}{x}  \Biggl[
            2 ( \zeta(2) + 2 ) H( - r, - r,x) 
          - 3 H( - r, - r,-1,x)
          - 2 H( - r,-4, - r,x) \nn\\
& & 
          + 6 H( - r, - r,-1,-1,x)
          - 3 H( - r, - r,0,-1,x)
          + H( - r,-4,-4, - r,x) \nn\\
& & 
          + \frac{3}{2} H( - r,-4, - r,-1,x)
          - H(0,0,-1,x)
          + 2 H(0,0,-1,-1,x) \nn\\
& & 
          - \frac{1}{2} H(0,0,0,-1,x)
          \Biggr] 
\, .
\eea

\be
\parbox{20mm}{\begin{fmfgraph*}(15,15)
\fmfleft{i}
\fmfright{o1,o2}
\fmfforce{0.2w,0.5h}{v1}
\fmf{dashes}{i,v1}
\fmf{photon}{v4,o1}
\fmf{photon}{v5,o2}
\fmf{plain,tension=.2}{v1,v3}
\fmf{photon,tension=.4}{v3,v5}
\fmf{plain,tension=.15}{v1,v4}
\fmf{photon,tension=0}{v4,v5}
\fmf{photon,tension=0}{v4,v3}
\end{fmfgraph*}}  =  \left( \frac{\mu^{2}}{a} \right) ^{2 \epsilon} 
\sum_{i=-1}^{0} \epsilon^{i} F^{(7)}_{i} + {\mathcal O} \left( 
\epsilon \right) , 
\ee
where:
\bea
aF^{(7)}_{-1} & = &   \frac{1}{2x}  \Bigl[
            3 H( - r, - r,-1;x)
          - 2 H(0, - r, - r;x)
          - H(0,0,-1;x)
		     \Bigr]
\, , \\
aF^{(7)}_{0} & = &   \frac{1}{2x}  \Bigl[
            6 H( - r, - r, - r, - r;x)
          - 12 H( - r, - r,-1,-1;x) \nn\\
& & 
          + 6 H( - r, - r,0,-1;x)
          - 3 H( - r,-4, - r,-1;x)
          + 2 H(0, - r,-4, - r;x) \nn\\
& & 
          - 2 H(0,0, - r, - r;x)
          + 4 H(0,0,-1,-1;x)
          - H(0,0,0,-1;x)
		     \Bigr]
 \, .
\eea

\vskip 1.2truecm
\boldmath
\centerline{\bf Topology $t=6$ \label{6den}}
\unboldmath
\vskip 0.8truecm

\be
\parbox{20mm}{\begin{fmfgraph*}(15,15)
\fmfleft{i}
\fmfright{o1,o2}
\fmfforce{0.2w,0.5h}{v1}
\fmf{dashes}{i,v1}
\fmf{photon}{v4,o1}
\fmf{photon}{v5,o2}
\fmf{plain,tension=.3}{v1,v2}
\fmf{plain,tension=.3}{v1,v3}
\fmf{photon,tension=.3}{v2,v4}
\fmf{photon,tension=.3}{v3,v5}
\fmf{photon,tension=0}{v2,v5}
\fmf{photon,tension=0}{v3,v4}
\end{fmfgraph*}}  = \left( \frac{\mu^{2}}{a} \right) ^{2 \epsilon} 
\sum_{i=-2}^{0} \epsilon^{i} F^{(8)}_{i} + {\mathcal O} \left( 
\epsilon \right) , 
\ee
where:
\bea
a^2F^{(8)}_{-2} & = &  \frac{6}{x\sqrt{x(x+4)}} H( - r,-1;x)
\, , \\
a^2F^{(8)}_{-1} & = &   \frac{1}{x\sqrt{x(x+4)}} \Bigl[
            16 H( - r, - r, - r;x) 
          - 24 H( - r,-1,-1;x) \nn\\
& & 
          + 16 H( - r,0,-1;x) 
          - 12 H(-4, - r,-1;x) 
		  \Bigr]
\, , \\
a^2F^{(8)}_{0} & = &   \frac{1}{x\sqrt{x(x+4)}} \Bigl[
          - 12 H( - r, - r, - r,-1;x)
          - 16 H( - r, - r,-4, - r;x) \nn\\
& & 
          + \! 12 \zeta(2) H( - r,\! -1;x)
          + \! 96 H( - r,\! -1,\! -1,\! -1;x)
          - \! 36 H( - r,\! -1,0,\! -1;x) \nn\\
& & 
          + \! 24 H( - r,0, - r, - r;x)
          - \! 64 H( - r,0,\! -1,\! -1;x)
          + \! 24 H( - r,0,0,\! -1;x) \nn\\
& & 
          - 32 H(-4, - r, - r, - r;x)
          + 48 H(-4, - r,-1,-1;x) \nn\\
& & 
          - 32 H(-4, - r,0,-1;x)
          + 24 H(-4,-4, - r,-1;x)
		  \Bigr]
\, .
\eea

\vskip 1truecm
\centerline{\bf Acknowledgments}
\vskip 0.6truecm

We are grateful to J.~Vermaseren for his kind assistance in the use
of the algebra manipulating program {\tt FORM}~\cite{FORM}, by which
all our calculations were carried out.

We wish to thank E.~Remiddi for discussions and for the {\tt C} 
program {\tt SOLVE} \cite{SOLVE}, used to solve the linear systems 
generated with the ibp identities.

\vskip 1.2truecm
\centerline{\bf A One-loop vertex}
\vskip 0.8truecm

The renormalization of the two-loop corrections to process (\ref{process2})
requires the knowledge of the one-loop vertex with three equal masses 
and of its derivative with respect to the squared mass $a$, 
which are given below.

\be
\parbox{20mm}{\begin{fmfgraph*}(15,15)
\fmfleft{i}
\fmfright{o1,o2}
\fmfforce{0.2w,0.5h}{v1}
\fmfforce{0.83w,0.13h}{v4}
\fmfforce{0.83w,0.87h}{v5}
\fmf{dashes}{i,v1}
\fmf{photon}{v4,o1}
\fmf{photon}{v5,o2}
\fmf{plain}{v1,v5}
\fmf{plain}{v1,v4}
\fmf{plain}{v4,v5}
\end{fmfgraph*}}  =  \left( \frac{\mu^{2}}{a} \right) ^{\epsilon} 
\sum_{i=0}^{2} \epsilon^{i} F^{(9)}_{i} + {\mathcal O} \left( 
\epsilon^{3} \right) , 
\ee
where:
\bea
aF^{(9)}_{0} & = &   \frac{1}{x} H(-r,-r;x)       \, , \\
aF^{(9)}_{1} & = & - \frac{1}{x} H(-r,-4,-r;x)    \, , \\
aF^{(9)}_{2} & = &   \frac{1}{x} H(-r,-4,-4,-r;x) \, .
\eea

\be
\frac{\partial}{\partial a} \, \, 
\parbox{20mm}{\begin{fmfgraph*}(15,15)
\fmfleft{i}
\fmfright{o1,o2}
\fmfforce{0.2w,0.5h}{v1}
\fmfforce{0.83w,0.13h}{v4}
\fmfforce{0.83w,0.87h}{v5}
\fmf{dashes}{i,v1}
\fmf{photon}{v4,o1}
\fmf{photon}{v5,o2}
\fmf{plain}{v1,v5}
\fmf{plain}{v1,v4}
\fmf{plain}{v4,v5}
\end{fmfgraph*}}  =  \left( \frac{\mu^{2}}{a} \right) ^{\epsilon} 
\sum_{i=0}^{2} \epsilon^{i} F^{(10)}_{i} + {\mathcal O} \left( 
\epsilon^{3} \right) , 
\ee
where:
\bea
a^2F^{(10)}_{0} & = & - \frac{1}{\sqrt{x(x+4)}} H(-r;x)       \, , \\
a^2F^{(10)}_{1} & = & - \frac{1}{x} H(-r,-r;x)  
                      + \frac{1}{\sqrt{x(x+4)}} H(-4,-r;x)    \, , \\
a^2F^{(10)}_{2} & = &   \frac{1}{x} H(-r,-4,-r;x) 
                      - \frac{1}{\sqrt{x(x+4)}} H(-4,,-4,-r;x) \, .
\eea
%
%
%
%
%

\end{fmffile}

\end{document}